\documentclass[12pt,a4paper]{article}
\usepackage[utf8]{inputenc}
\usepackage[T1]{fontenc}
\usepackage[american]{babel}
\usepackage{hyperref}
\usepackage{graphicx,color}
\usepackage{relsize}
\usepackage{cancel}
\usepackage{amsmath,amssymb,amsfonts}
\usepackage{bm}
\usepackage{tikz}
\usepackage{graphicx}
\title{Pair creation amplitudes for a real scalar field 
coupled to a time-dependent surface in $d+1$ dimensions}
\author{C.~D.~Fosco and B.~C.~Guntsche \\
{\normalsize\it Centro At\'omico Bariloche and Instituto Balseiro}\\
{\normalsize\it Comisi\'on Nacional de Energ\'\i a At\'omica}\\
{\normalsize\it R8402AGP Bariloche, Argentina.}}
\begin{document}
\date{}
\maketitle
%====================================================================
\begin{abstract}
We study the pair creation phenomenon for a
real scalar field $\varphi$ in the presence of a surface that undergoes
time-dependent deformations, while imposing Dirichlet-like boundary
conditions. Including terms up to fourth order in the departure of the 
surface from an infinite plane, we present results for the 
angular dependence of the emission rate for the vacuum-to-pair process as a 
function of the geometry and the dynamics of the surface, as well as of the 
momenta of the emitted pair. 
We check the consistency of the leading contribution with previous results
obtained from the imaginary part of the effective action, and clarify how the
relation between exclusive probabilities and the imaginary part of the effective action is modified at
fourth order by the opening of a two-pair channel.
\end{abstract}
%%%%%%%%%%%%%%%%%%%%%%%%%%%%%%%%%%%%%%%%%%%%%%%%%%%%%%%%%%%%%%%%%%%%%%%%%%%
%%%%%%%%%%%%%%%%%%%%%%%%%%%%%%%%%%%%%%%%%%%%%%%%%%%%%%%%%%%%%%%%%%%%%%%%%%% 
%%%%%%%%%%%%%%%%%%%%%%%%%%%%% Introduction %%%%%%%%%%%%%%%%%%%%%%%%%%%%%%%%
%%%%%%%%%%%%%%%%%%%%%%%%%%%%%%%%%%%%%%%%%%%%%%%%%%%%%%%%%%%%%%%%%%%%%%%%%%%
%%%%%%%%%%%%%%%%%%%%%%%%%%%%%%%%%%%%%%%%%%%%%%%%%%%%%%%%%%%%%%%%%%%%%%%%%%%
\section{Introduction}\label{sec:intro}
The Dynamical Casimir Effect (DCE), whereby particles are created out of the
quantum vacuum in some time-dependent environments, typically boundary
conditions, has been extensively studied in the
literature~\cite{moore1970,FulDav,reviews,milton1999casimir,
bordag2009advances}.
In a companion paper~\cite{FG2024}, we have evaluated the imaginary part of the
effective action $\Gamma(\psi)$ for a massless real scalar field 
$\varphi$ in $d+1$ dimensions, subjected to Dirichlet boundary conditions on a 
time-dependent surface~$\Sigma$. Since the imaginary part of the real-time 
effective action determines the total probability of vacuum 
decay~\cite{Schwinger1951}, that 
calculation yields the \emph{inclusive} probabilities of pair creation, summed 
over all possible final states.

In this work, we adopt a complementary viewpoint, aimed at computing the
\emph{exclusive} transition amplitude from the vacuum to a specific two-particle
final state. To that end, we use standard quantum field theory perturbation 
theory in the interaction representation, with the ``free'' part of the system
being defined as one where the field sees a static and planar surface. In 
order to make the calculation well-defined from the mathematical point of view, 
rather than beginning from a surface which imposes Dirichlet conditions, we
use a slightly more general setup: Dirichlet conditions are recovered in the 
limit in which the parameter $\lambda$, that characterizes the surface tends 
to infinity; that limit is taken at the end of the calculation.

We explore rather general time-dependent surfaces and include 
higher-order terms. We wish to point out that the angular dependence 
of the emitted radiation for a rigidly moving mirror coupled to the 
electromagnetic field has been considered in~\cite{MaiaNeto1996}, a 
configuration for which we present the scalar field equivalent 
here, as an example. 

The general expressions apply to rather general time-dependent surfaces; in 
particular, we analyze a Dirichlet surface that is flat except for a small, 
harmonically oscillating bump, and show that the radiation follows a Lambert 
pattern. We also include higher-order terms and relate them to our previous 
result for the imaginary part of the effective action in the same system.

This paper is organized as follows. In Sec.~\ref{sec:thesystem}, we define the
system and construct the perturbative expansion. 
In Sec.~\ref{sec:results} we present explicit results for the first 
nontrivial orders as well as applications to nontrivial examples. 
In Sec.~\ref{sec:consistency}, we verify the consistency of our results with 
the imaginary part of the effective action computed in~\cite{FG2024}. Finally, 
in Sec.~\ref{sec:conc} we present our conclusions.
%%%%%%%%%%%%%%%%%%%%%%%%%%%%%%%%%%%%%%%%%%%%%%%%%%%%%%%%%%%%%%%%%%%%%%%%%%%
%%%%%%%%%%%%%%%%%%%%%%%%%%%%%%%%%%%%%%%%%%%%%%%%%%%%%%%%%%%%%%%%%%%%%%%%%%%
%%%%%%%%%%%%%%%%%%%%%%%%%%%%%% The system %%%%%%%%%%%%%%%%%%%%%%%%%%%%%%%%%
%%%%%%%%%%%%%%%%%%%%%%%%%%%%%%%%%%%%%%%%%%%%%%%%%%%%%%%%%%%%%%%%%%%%%%%%%%%
%%%%%%%%%%%%%%%%%%%%%%%%%%%%%%%%%%%%%%%%%%%%%%%%%%%%%%%%%%%%%%%%%%%%%%%%%%%
\section{The system}\label{sec:thesystem}
\subsection{Definitions and conventions}
We consider a massless real scalar field $\varphi(x)$ in $d+1$ 
space-time dimensions. To describe the system, we adopt the convention of using 
the Minkowski 
metric $g_{\mu\nu} = \mathrm{diag}(+1,-1,-1,\ldots,-1)$.
Space-time indices $\mu,\nu,\ldots$ run over the values $0,1,\ldots,d$,
with $x^0 = t$ denoting the real time. We also use natural 
units, $\hbar = c = 1$, and Einstein's summation convention.

The scalar field interacts with a space-time surface $\Sigma$, which we assume  
to be described by a single Monge patch; namely, letting $x_\shortparallel 
\equiv (x^\alpha)_{\alpha=0}^{d-1}$ denote the first $d$ space-time
coordinates, the surface is parametrized as
\begin{equation}\label{eq:defmonge}
\Sigma\;:\; x_d \,=\, \psi(x_\shortparallel) \;,
\end{equation}
where $\psi(x_\shortparallel)$ specifies the ``height'' of the surface, measured
with respect to the hyperplane $x_d = 0$.
The action is assumed to have the form:
\begin{equation}\label{eq:defs}
{\mathcal S}(\varphi;\psi) \,=\, \frac{1}{2} \int d^{d+1}x \,\Big[
\partial_\mu \varphi(x) \, \partial^\mu \varphi(x)
- \lambda\, \delta\big(x^d - \psi( x_\shortparallel) \big) 
\big( \varphi(x) \big)^2 
\Big]
\;.
\end{equation}
 The constant $\lambda >0$ has the dimensions of a mass, and plays the role of 
 penalizing the development of nonvanishing values for the field $\varphi$ on 
 $\Sigma$. Dirichlet conditions: $\varphi\big(x_\shortparallel , \, 
 \psi(x_\shortparallel)\big)=0$ appear when $\lambda \to +\infty$ at the end 
 of the calculation. In this vein, since $\lambda$ plays an auxiliary role and 
 in general we do not attempt here to obtain final results for finite 
 $\lambda$, except in a $d=1$ example, where the simplicity of the final 
 expressions warrants their presentation.

With the aim of finding the probability distributions of the pairs created
by the moving surface, we shall evaluate the transition amplitudes using perturbation theory in the interaction representation. Here we take 
as the perturbation the deformation of the surface, which in our setup is 
characterized by the single function $\psi$.  
Thus, we split up the action  ${\mathcal S}$  into its free (${\mathcal 
S}_0$) and 
interaction (${\mathcal S}_I$) 
parts:
\begin{equation}
{\mathcal S}(\varphi;\psi) \;=\; 
{\mathcal S}_0(\varphi) \,+\,
{\mathcal S}_I(\varphi;\psi) \;,
\end{equation}
such that:
\begin{equation}\label{eq:defs0}
{\mathcal S}_0(\varphi) \,=\, \frac{1}{2} \int d^{d+1}x \,\Big[
\partial_\mu \varphi(x) \, \partial^\mu \varphi(x)
- \lambda\, \delta(x^d) \,\big( \varphi(x) \big)^2 \Big]
\;,
\end{equation}
and
\begin{align}\label{eq:defsi}
{\mathcal S}_I(\varphi;\psi) \,=\, - \frac{\lambda}{2} \int d^{d+1}x \,\Big[
\delta\big(x^d - \psi( x_\shortparallel) \big) - \delta\big(x^d\big)\Big]
	\big( \varphi(x) \big)^2 \;.
\end{align}
In the interaction representation~\footnote{No additional notation will be 
introduced, for the sake of clarity, to denote the use of this 
representation.}, the field evolves according to the dynamics dictated by 
${\mathcal S}_0$. 

\subsection{Free field states and Feynman propagator}
The space of states into which the vacuum $|0\rangle$ may decay is 
built by studying the solutions to the free equation of motion,
$[\Box + \lambda \delta(x^d)] \varphi(x) = 0$.   
A convenient way to construct such a set is by taking advantage of the
translation invariance along ${\mathbf x}_\shortparallel \equiv (x^1, 
\ldots \, x^{d-1})$. Namely, introducing the momentum, $k_\shortparallel \equiv 
(k^0, {\mathbf k_\shortparallel})$, a convenient basis of solutions is then 
given by 
the following functions:
\begin{equation}
g^{(o,e)}_k(x) \, = \, {\mathcal N}_k^{(o,e)} \, e^{-i k_\shortparallel \cdot 
x_\shortparallel } \, f_{k^d}^{(o,e)}(x^d) 
\end{equation}
where ${\mathcal N}_k^{(o,e)}$ are normalization constants and 
$k_\shortparallel \cdot x_\shortparallel \equiv \sum_{\alpha=0}^{d-1} k_\alpha 
x^\alpha$, satisfies $k_\shortparallel^2 = (k^d)^2$ (on-shell condition).
Finally, the $f_{k^d}^{(o,e)}$ functions denote the (odd and even, 
respectively) functions:
\begin{align}\label{eq:feo}
	f_{k^d}^{(o)}(x^d) &=\; \sin(k^d x^d) \nonumber\\
	f_{k^d}^{(e)}(x^d) &=\; \sin(k^d |x^d|) \,+\, 
	\frac{2 k^d}{\lambda} \cos(k^d x^d) \;,
\end{align}
defined for $|x^d| \leq \frac{L}{2}$, $L$ being the side length of a 
spatial box. As usual, this is introduced to avoid the
infinities associated with a continuum spectrum~\footnote{On the other 
hand, any reference to the finite size shall be eliminated at the end of the 
calculation, when evaluating probability densities.}. Indeed, assuming
periodic boundary conditions at $x^d = \pm \frac{L}{2}$, the allowed values of 
the momentum become: \mbox{$k^d = \frac{2 n 
\pi}{L}$}, $n \in {\mathbb N}$, for the odd and even functions.
On the other hand, using the same $L$ for the other directions: ${\mathbf 
k_\shortparallel} = \frac{2 \pi}{L} \, {\mathbf n_\shortparallel}$, ${\mathbf 
n_\shortparallel} \in {\mathbb Z}^{(d-1)}$. 

The normalization factors become 
\begin{equation}
{\mathcal N}_k^{(o)} \,=\, \Big(L^d |k^0| \Big)^{-\frac{1}{2}} 
\;,\;\;
{\mathcal N}_k^{(e)} \,=\, \Big\{ L^d |k^0|
\big[ 1 + (\frac{2 k^d}{\lambda} )^2\big] \Big\}^{-\frac{1}{2}} \;,
\end{equation}
and the resulting free field decomposition can be put in the form:
\begin{equation}
\varphi(x) \;=\; \sum_{{\mathbf n_\shortparallel} \in {\mathbb Z}^{d-1} , n^d 
\in 
{\mathbb N}}
\sum_{\alpha=o,e} \Big[ a_{\mathbf k}^{(\alpha)} \, g_{\mathbf 
k}^{(\alpha)}(x) 
\,+\,
(a_{\mathbf k}^{(\alpha)})^\dagger \, (g_{\mathbf k}^{(\alpha)})^*(x) 
\Big] \,
\end{equation}
where ${\mathbf k} \equiv \frac{2 \pi}{L} (n^1,\ldots,n^d)
\equiv \frac{2 \pi}{L} {\mathbf n}$.

With the normalization we have chosen we have: 
$\big(g_{\mathbf 
k}^{(\alpha)},g_{\mathbf k'}^{(\alpha')}\big) = \delta_{\alpha \alpha'} 
\delta_{{\mathbf n} {\mathbf n}'}$, with
\begin{equation}\label{eq:defsp}
	\big(f , g \big) \,\equiv \, i \int_{\mathbf x}  
	f^*(x) \stackrel{\leftrightarrow}{\partial_0} g(x) \;,
\end{equation}
and as a consequence $[a_{\mathbf k}^{(\alpha)} , a_{\mathbf 
k'}^{(\alpha')\dagger}] = \delta_{\alpha \alpha'}  \delta_{{\mathbf n} {\mathbf 
n}'}$.  We have used in (\ref{eq:defsp}) a shorthand notation for the integral, 
whereby rather than writing differentials, the integrated variables appear 
as sub-indices for the integral symbol. In the case above, the range 
for that integration is the spatial hypercube of side $L$.

On the other hand, in order to evaluate the scattering amplitudes via Feynman 
rules, we also need the free Feynman propagator, 
\begin{equation}
G(x,x') \,\equiv\, \langle 0 | {\mathrm T}[\varphi(x) \varphi(x')] | 0 \rangle 
\;, 
\end{equation}
which satisfies the equation 
\begin{equation}
\big[- \Box_x - \lambda \delta(x^d) \big] G(x,x') \,=\, i 
\delta^{d+1}(x-x') \;.
\end{equation}
Fourier transforming with respect to the $x_\shortparallel$ coordinates,
\begin{equation}
G(x,x') \,\equiv\,\int \frac{d^dp_\shortparallel}{(2\pi)^d} e^{-i 
p_\shortparallel \cdot (x_\shortparallel - x_\shortparallel')} 
\widetilde{G}_{p_\shortparallel}(x^d,x'^d) \;,
\end{equation}
the problem becomes one-dimensional, and we find:
\begin{equation}
\widetilde{G}_{p_\shortparallel}(x^d,x'^d) \,=\,
\frac{1}{2 i \kappa(p_\shortparallel)}
\left[ e^{- \kappa(p_\shortparallel) |x^d - x'^d|}
\,-\, r(p_\shortparallel) \, 
e^{- \kappa(p_\shortparallel) (|x^d| + |x'^d|)} \right] \;,
\end{equation}
where $\kappa(p_\shortparallel) \equiv \sqrt{- p_\shortparallel^2 - i \epsilon}$ 
and
$r(p_\shortparallel) = \frac{\lambda}{\lambda + 2 \kappa(p_\shortparallel)}$. 

\subsection{Emission Probabilities from the Perturbative Amplitudes}
We denote by ${\mathcal A}\big({\mathbf k}, \alpha ; {\mathbf k'}, \alpha' 
\big)$ the amplitude from the vacuum $|0\rangle$ to a \mbox{$2$-particle} 
state $|{\mathbf k} \alpha , {\mathbf k'} \alpha' \rangle$ which, in the 
interaction representation, is obtained as follows:
\begin{equation}\label{eq:Adef}
{\mathcal A}\big({\mathbf k}, \alpha ; {\mathbf k'}, \alpha' \big) 
\,=\,
\langle {\mathbf k} \alpha , {\mathbf k'} \alpha' | 
T \big( e^{i :{\mathcal S}_I: } \big) 
|0\rangle \;,
\end{equation}
$T$ being the time-ordering operator and $:{\mathcal S}_I:$ the normal 
ordered version of (\ref{eq:defsi}).  

A clarification about (\ref{eq:Adef}) will be important when, in 
Sec.~\ref{sec:consistency}, we relate these amplitudes to the imaginary part 
of the effective action. When the right-hand side is evaluated perturbatively 
through Wick's theorem, each contraction factorizes into a piece attached to 
the two external legs and a set of disconnected vacuum subdiagrams; the latter 
exponentiate into the vacuum persistence amplitude 
${\mathcal V}\equiv\langle 0|S|0\rangle$, so that 
$\langle {\mathbf k}\alpha,{\mathbf k'}\alpha'|S|0\rangle = {\mathcal V}\,
{\mathcal A}({\mathbf k},\alpha;{\mathbf k'},\alpha')$. Throughout this work 
${\mathcal A}$ denotes the \emph{connected} amplitude: the vacuum bubbles are 
factored out, which amounts to dividing by ${\mathcal V}$,
\begin{equation}\label{eq:Aconn}
{\mathcal A}\big({\mathbf k},\alpha;{\mathbf k'},\alpha'\big)\,=\,
\frac{\langle {\mathbf k}\alpha,{\mathbf k'}\alpha'|S|0\rangle}
{\langle 0|S|0\rangle}\;.
\end{equation}
For a \emph{stable} vacuum ${\mathcal V}$ is a pure phase and this division is 
immaterial for any probability; the moving surface, however, destabilizes the 
vacuum and $|{\mathcal V}|^2 = e^{-2\,{\rm Im}\,\Gamma}<1$ is no longer trivial. 
Consequently $\tfrac12|{\mathcal A}|^2$ in (\ref{eq:defprob}) is the 
\emph{connected} probability, which coincides with the true single-pair 
probability only at leading order; the factor $|{\mathcal V}|^2$ is not lost but 
reinstated in Sec.~\ref{sec:consistency} [see (\ref{eq:squeezed}) and the 
probabilities $p_1,p_2$ that follow it], where it encodes the depletion of the 
vacuum into the multi-pair channels that open at fourth order.

The probabilities of detecting particles as functions of their momenta and of 
the deformation of the surface result from taking the modulus 
squared of the amplitudes. The latter will be evaluated as power 
series in $\psi$ and, to get probabilities up to fourth order, 
we only need the amplitudes up to third order in $\psi$. Indeed, the 
probability for a particle pair with momenta ${\mathbf k}$ and 
${\mathbf k'}$, within infinitesimal volume elements $d^d{\mathbf k}$  
$d^d{\mathbf k'}$ is:
\begin{equation}\label{eq:defprob}
dP\big({\mathbf k} , {\mathbf k'}\big) \,=\, \frac{1}{2} \,
\big|{\mathcal A}\big({\mathbf k} ; {\mathbf k'}\big)\big|^2 
\,\frac{ L^d d^d{\mathbf k}}{(2 \pi)^d} \, 
\,\frac{L^d d^d{\mathbf k'} }{(2\pi)^d} \,,
\end{equation}
where the factor $\frac{1}{2}$ is the usual Bose-statistics factor for final 
states of identical particles, and we have used the shorthand
\begin{equation}\label{eq:Asquared}
\big|{\mathcal A}\big({\mathbf k} ; {\mathbf k'}\big)\big|^2 \,\equiv\,
\sum_{\alpha,\alpha' = o,e} \,
	\big|{\mathcal A}\big({\mathbf k},\alpha;\, 
	{\mathbf k'},\alpha'\big)\big|^2
\end{equation}
for the incoherent sum over the parities ($e$, $o$) of the emitted 
quanta, states which are mutually orthogonal. Also note that the powers of $L$ will cancel with the ones coming from the 
normalization factors of the final states.

To understand the relationship between the expansion of amplitudes and that of 
the probabilities, we expand ${\mathcal A}$ in powers of 
$\psi$,
\begin{equation}
{\mathcal A}\big({\mathbf k},\alpha;\, 
{\mathbf k'},\alpha'\big) \,=\,{\mathcal A}^{(1)}\big({\mathbf k},\alpha;\, 
{\mathbf k'},\alpha'\big)+{\mathcal A}^{(2)}\big({\mathbf k},\alpha;\, 
{\mathbf k'},\alpha'\big)+{\mathcal A}^{(3)}\big({\mathbf k},\alpha;\, 
{\mathbf k'},\alpha'\big)+\ldots
\end{equation}

To find the probabilities, one needs to consider: 
\begin{align}
\big|{\mathcal A}\big({\mathbf k},\alpha;\, {\mathbf k'},\alpha'\big)\big|^2 
& =\,\big|	{\mathcal A}^{(1)}\big({\mathbf k},\alpha;\, 
{\mathbf k'},\alpha'\big) \big|^2 \nonumber\\	
& + \, 2\, {\mathrm Re}\Big[{\mathcal A}^{(1) *}\big({\mathbf k},\alpha;\, 
{\mathbf k'},\alpha'\big) {\mathcal A}^{(2)}\big({\mathbf k},\alpha;\, 
{\mathbf k'},\alpha'\big) \Big] 
 \nonumber\\
& +\,\big|{\mathcal A}^{(2)}\big({\mathbf k},\alpha;\, 
{\mathbf k'},\alpha'\big)\big|^2 \,+\, 2\, {\mathrm Re}\Big[{\mathcal A}^{(1) 
*}\big({\mathbf k},\alpha;\, 
{\mathbf k'},\alpha'\big) {\mathcal A}^{(3)}\big({\mathbf k},\alpha;\, 
{\mathbf k'},\alpha'\big) \Big]\nonumber\\
&+\, \ldots
\end{align}
where, on the right-hand side, the first line contributes to the probability at second order 
in $\psi$, the second line at third order, etc. It is therefore clear 
that the probability to fourth order requires the amplitude only up to 
third order. Moreover, a subtler point is noteworthy, and applies to terms 
involving products of amplitudes of different orders (like the third-order 
one, which involves the second- and first-order amplitudes).
The point is that the factors in one such product have the same quantum 
numbers, in particular the parity. Thus, if the factors have different selection
rules regarding parity, the product vanishes. This happens, as we shall see
for the first- and second-order amplitudes.

\subsection{Perturbative Expansion for the  Amplitudes} \label{subsec:pertamp}
Let us then calculate each ${\mathcal A}^{(i)}$, which requires us to 
expand both the interaction action and its exponential.

We first note that, using an index to denote the respective order in $\psi$,
we have 
${\mathcal S}_I =
{\mathcal S}_I^{(1)} + {\mathcal S}_I^{(2)} \,+\,\ldots$, with
\begin{align}
{\mathcal S}_I^{(1)} &=\,-\lambda \int d^d x_\shortparallel \, 
\psi( x_\shortparallel) \, \varphi(x_\shortparallel, 0) 
\partial_d\varphi(x_\shortparallel, 0) \nonumber\\
{\mathcal S}_I^{(2)} &=\, - \frac{\lambda}{2} 
\int d^d x_\shortparallel \, \big(\psi( x_\shortparallel) \big)^2 \, 
[ \partial_d\varphi(x_\shortparallel, 0) \partial_d\varphi(x_\shortparallel, 0)
+  \varphi(x_\shortparallel, 0) \partial^2_d\varphi(x_\shortparallel, 
0)]\nonumber\\
{\mathcal S}_I^{(3)} &=\, - \frac{\lambda}{6}\, 
\int d^d x_\shortparallel \, \big(\psi( x_\shortparallel) \big)^3 \, 
[ 3 \, \partial_d\varphi(x_\shortparallel, 0) 
\partial^2_d\varphi(x_\shortparallel, 0)
+  \varphi(x_\shortparallel, 0) \partial^3_d\varphi(x_\shortparallel, 0)] 
\nonumber\\
\ldots
\end{align} 

The amplitudes at each order are obtained by combining different powers of the interaction, and suitable orders in the
expansion of the interaction itself.

The only case in which one does not need to combine those two expansions is the 
first-order term. Introducing Fourier transforms according to the 
convention
$
\widetilde{\psi}(p_\shortparallel) =\int d^dx_\shortparallel\, e^{i 
p_\shortparallel \cdot x_\shortparallel} \,\psi(x_\shortparallel)
$, the first-order amplitude becomes, for any $\lambda$:
\begin{align}\label{eq:defa1}
	{\mathcal A}^{(1)}\big({\mathbf k}, \alpha ; 
	{\mathbf k'}, \alpha' \big) \,=\, -i \lambda \, {\mathcal N}_k^{(\alpha)} 
	{\mathcal N}_{k'}^{(\alpha')} \,
	\widetilde{\psi}(k_\shortparallel + {k'}_\shortparallel) \,& 
	\Big[ \big(f_{k^d}^{(\alpha)} \partial_d 
	f_{{k'}^d}^{(\alpha')}\big)|_{x^d=0}
	\nonumber\\
	& + (k^d,\alpha) \leftrightarrow ({k'}^d,\alpha')\big] \;.
\end{align}
We note that, for the factor between square brackets to be nonvanishing, the 
parities under $x^d \to - x^d$ of the emitted particles, $\alpha$ and 
$\alpha'$, have to be different: one odd and the other even. 
More explicitly,
\begin{align}\label{eq:eqa1}
	{\mathcal A}^{(1)}\big({\mathbf k}, e ; 
	{\mathbf k'}, o \big) & =\, - 2 i \, {\mathcal N}_k^{(e)} 
	{\mathcal N}_{k'}^{(o)} \,
	\widetilde{\psi}(k_\shortparallel + {k'}_\shortparallel) k^d {k'}^d 
	\nonumber\\
&= {\mathcal A}^{(1)}\big({\mathbf k'}, o ; 
	{\mathbf k}, e \big) \;, \nonumber\\
{\mathcal A}^{(1)}\big({\mathbf k}, e ; {\mathbf k'}, e \big) &=\,
 {\mathcal A}^{(1)}\big({\mathbf k}, o ; {\mathbf k'}, o \big) = 0\;.
\end{align}

The second-order term consists of two contributions: the first comes from
the second-order term in the expansion of ${\mathcal S}_I$; the second 
comes from the square of the first-order term. Important 
cancellations occur between those two terms. We simply write down the sum 
of them, in the $\lambda \to \infty$ limit:
\begin{align}
{\mathcal A}^{(2)}\big({\mathbf k}, \alpha ; 
{\mathbf k'}, \alpha' \big) \,=\, 
{\mathcal A}^{(2,1)}\big({\mathbf k}, \alpha ; {\mathbf k'}, \alpha' \big) 
\,+\, 
{\mathcal A}^{(2,2)}\big({\mathbf k}, \alpha ; {\mathbf k'}, \alpha' \big) 
\nonumber\\
=\, -2 i \, \big({\mathcal N}_k\big)^2 
\;
\int \frac{d^d p_\shortparallel}{(2\pi)^d} \,
\widetilde{\psi}(k_\shortparallel - p_\shortparallel) \,
\sqrt{- p_\shortparallel^2} \,
\widetilde{\psi}(p_\shortparallel + k'_\shortparallel) \, k^d {k'}^d
\, \delta^{\alpha \alpha'} \;,
\end{align}
(no sum over $\alpha$, $\alpha'$) where $
{\mathcal N}_k = {\mathcal N}_k^{(\alpha)}\Big|_{\lambda \to \infty}
= \Big(L^d |k^0| \Big)^{-\frac{1}{2}}$.

It then becomes evident, without further ado, that the third-order contribution 
to the decay probability vanishes. Indeed, it involves the product of 
amplitudes to first and second orders, with the same initial and final 
states. However, the first-order term requires the parities to be different, while 
the second-order one vanishes unless they are equal.

Finally, the third-order amplitude is nonvanishing only when the parities are 
different, as was the case for the first-order one. We therefore write it 
directly for one of the nontrivial cases, and also in the 
$\lambda \to \infty$ limit:
\begin{align}
{\mathcal A}^{(3)}\big({\mathbf k}, e ; 
{\mathbf k'}, o \big) &=\, i \,  \big({\mathcal N}_k\big)^2  \, 
k^d {k'}^d \, \Big\{ \frac{1}{3} \,
\widetilde{\psi^3}(k_\shortparallel + k'_\shortparallel) \, (k'^d)^2 
\nonumber\\
& +\,\int \frac{d^d p_\shortparallel}{(2\pi)^d} \,
\widetilde{\psi}(k'_\shortparallel - p_\shortparallel) \,
p_\shortparallel^2 \,\widetilde{\psi^2}(p_\shortparallel + k_\shortparallel) \,
\nonumber\\
 - \, 2  \,\int \frac{d^d p_\shortparallel}{(2\pi)^d} 
 \frac{d^d q_\shortparallel}{(2\pi)^d} &
\widetilde{\psi}(k_\shortparallel - p_\shortparallel) \,
\sqrt{- p_\shortparallel^2} \,
\widetilde{\psi}(p_\shortparallel - q_\shortparallel) \,
\sqrt{- q_\shortparallel^2} 
\,
\widetilde{\psi}(q_\shortparallel + k'_\shortparallel) 
\Big\}	\,. 
\end{align}

%%%%%%%%%%%%%%%%%%%%%%%%%%%%%%%%%%%%%%%%%%%%%%%%%%%%%%%%%%%%%%%%%%%%%%%%%%%
%%%%%%%%%%%%%%%%%%%%%%%%%%%%%%%%%%%%%%%%%%%%%%%%%%%%%%%%%%%%%%%%%%%%%%%%%%%
%%%%%%%%%%%%%%%%%%%%%%%%%%%%%%% Results %%%%%%%%%%%%%%%%%%%%%%%%%%%%%%%%%%%
%%%%%%%%%%%%%%%%%%%%%%%%%%%%%%%%%%%%%%%%%%%%%%%%%%%%%%%%%%%%%%%%%%%%%%%%%%%
%%%%%%%%%%%%%%%%%%%%%%%%%%%%%%%%%%%%%%%%%%%%%%%%%%%%%%%%%%%%%%%%%%%%%%%%%%%
\section{Results and Applications}\label{sec:results}
\subsection{Results}
The second-order probabilities are obtained from the first-order 
amplitudes. Recalling  (\ref{eq:defa1}), 
\begin{align}
{\mathcal A}^{(1)}\big({\mathbf k}, e ; {\mathbf k'}, o \big) & =\, -2 i 
L^{-d}\, 
\frac{\widetilde{\psi}(k_\shortparallel + {k'}_\shortparallel) \,
	k^d {k'}^d}{ \big( |k^0| |{k'}^0|\big)^{\frac{1}{2}} } \,
	\big[ 1 + (\frac{2 {k}^d}{\lambda} 
)^2\big]^{-\frac{1}{2}} \;,
\end{align}
the probability of pair creation at second order in $\psi$, within a 
volume $d^d{\mathbf k} \, d^d{\mathbf k'}$ around ${\mathbf k}, {\mathbf k'}$, 
is
\begin{align}
	dP^{(2)}\big({\mathbf k} , {\mathbf k'}\big) 
	&=\, \frac{1}{2} \,\left[
	\big|{\mathcal A}^{(1)}\big({\mathbf k}, e ; {\mathbf 
	k'}, o\big)\big|^2 \, + \,
	\big|{\mathcal A}^{(1)}\big({\mathbf k}, o ; {\mathbf k'}, e\big)
	\big|^2 \right] \,\frac{ L^d d^d{\mathbf k}}{(2 \pi)^d} \, 
	\frac{L^d d^d{\mathbf k'}}{(2\pi)^d}  \nonumber\\ 
 = \, 
	\big|\widetilde{\psi}(k_\shortparallel + {k'}_\shortparallel)\big|^2  
	\, &
	\frac{2 |k^d|^2 |{k'}^d|^2}{|k^0| |{k'}^0|} 
	\Big\{\big[ 1 + (\frac{2 
		{k}^d}{\lambda} 
	)^2\big]^{-1} + \big[ 1 + (\frac{2 {k'}^d}{\lambda} 
	)^2\big]^{-1} \Big\} 	
	\frac{d^d{\mathbf k}}{(2\pi)^d}\, \frac{d^d{\mathbf k'}}{(2\pi)^d} \;.
\end{align}

Note that the $d^{\rm th}$ component of $k^\mu$ can assume only 
positive values, but the corresponding free particle states are not
plane waves. Indeed, they are mixtures of plane waves having 
positive and negative momenta.  All the other components run over positive and 
negative values. 

It is possible to obtain a more covariant looking representation for the 
probability:
\begin{align}
    dP^{(2)}\big({\mathbf k} , {\mathbf k'}\big) & =\, 8
\, \big|\widetilde{\psi}(k_\shortparallel + {k'}_\shortparallel)\big|^2  \, 
|k^d|^2 |{k'}^d|^2  \, \Theta(k^0) \delta(k^2) \, 
\Theta({k'}^0) \delta({k'}^2) \nonumber\\
   \times & \Big\{ \big[ 1 + 
(\frac{2 {k}^d}{\lambda})^2\big]^{-1} + \big[ 1 + (\frac{2 
{k'}^d}{\lambda} )^2\big]^{-1} \Big\}	
\frac{d^{d+1}k}{(2\pi)^d}\, 	\frac{d^{d+1}k'}{(2\pi)^d} \,,
\end{align}
where $\Theta$ denotes Heaviside's step function. 
This gives the equivalent form
\begin{align}\label{eq:dPcov2d}
	dP^{(2)}\big({\mathbf k}, {\mathbf k'} \big) & =\,2
	\, \big|\widetilde{\psi}(k_\shortparallel + {k'}_\shortparallel)\big|^2  \; 
	\theta(k^0) \,	\sqrt{k_\shortparallel^2} \; 	
	\Theta({k'}^{0}) \, \sqrt{{k'}_\shortparallel^2}  \nonumber\\
\times	& \Big\{\big[ 1 + \frac{4 
k_\shortparallel^2}{\lambda^2}\big]^{-1} +  
 \big[ 1 + \frac{4 {k'_\shortparallel}^2}{\lambda^2}\big]^{-1}
\Big\} 	
	\frac{d^dk_\shortparallel}{(2\pi)^d}\, 	
	\frac{d^d{k'}_\shortparallel}{(2\pi)^d} \,,
\end{align}
where, on the left-hand side, $k^d$ and ${k'}^d$ are assumed to be determined 
by the respective mass shell conditions: $k^d = \sqrt{k_\shortparallel^2}$ and
${k'}^d = \sqrt{{k'_\shortparallel}^2}$.
In the Dirichlet limit this becomes:
\begin{equation}\label{eq:dPgenD}
	dP^{(2)}\big({\mathbf k},\, {\mathbf k'} \big)\,=\, 4\,
	\big|\widetilde{\psi}(k_\shortparallel+{k'}_\shortparallel)\big|^2\,
	\Theta(k^0)\sqrt{k_\shortparallel^2}\,
	\Theta({k'}^{0})\sqrt{{k'}_\shortparallel^2}\,
	\frac{d^d k_\shortparallel}{(2\pi)^d}\,
	\frac{d^d {k'}_\shortparallel}{(2\pi)^d}\;.
\end{equation}

A useful expression, at this order, for the probability 
of detecting a particle at a given $k^\mu$ is the following:
\begin{equation}\label{eq:defp2}
	dP^{(2)}({\mathbf k}) \,=\, 4 \,\frac{d^d{\mathbf k}}{(2\pi)^d}
	\,	\frac{|k^d|^2}{|k^0|} \, \int \frac{d^d{k'}_\shortparallel}{(2\pi)^d}  
	\Theta(k'^0) \, \big|\widetilde{\psi}(k_\shortparallel + 
	{k'}_\shortparallel)\big|^2  \, 
	|{k'}_\shortparallel| \;.
\end{equation}

As already mentioned, the third-order probability for the
vacuum to decay to a pair vanishes. 
Indeed, it is evident that the third-order contributions will involve 
the product of a second-order amplitude and a first-order one; namely,
\begin{equation}
	{\mathcal A}^{(2)}\big({\mathbf k}, \alpha ; {\mathbf k'}, \alpha' \big)\,
	{\mathcal A}^{(1)*}\big({\mathbf k}, \alpha ; {\mathbf k'}, \alpha' \big)
\end{equation}
and its complex conjugate. But, as we have mentioned in the previous section, 
the first-order term is nonzero only when $\alpha \neq \alpha'$, 
while the second-order one only appears when $\alpha = \alpha'$. 
Thus, the probability density, to this order, vanishes.

Finally, we consider the fourth-order contribution,
which is obtained by collecting all the $O(\psi^4)$ pieces of
$\tfrac12\sum_{\alpha\alpha'}|{\mathcal A}|^2$. Using the parity selection rules
[${\mathcal A}^{(1)},{\mathcal A}^{(3)}$ require $\alpha\neq\alpha'$, while
${\mathcal A}^{(2)}$ requires $\alpha=\alpha'$, cf.\ Sec.~\ref{subsec:pertamp}], the only
surviving terms are
\begin{align}\label{eq:dP4master}
	dP^{(4)}\big({\mathbf k} , {\mathbf k'}\big)
	&=\,
	\Big\{
	\tfrac12\big|{\mathcal A}^{(2)}\big({\mathbf k}, e ; {\mathbf k'}, 
	e\big)\big|^2
	\,+\,
	\tfrac12\big|{\mathcal A}^{(2)}\big({\mathbf k}, o ; {\mathbf k'}, 
	o\big)\big|^2
	\nonumber\\
	&+\, {\mathrm Re}\Big[{\mathcal A}^{(1) *}\big({\mathbf k},e;
	{\mathbf k'},o\big)\, {\mathcal A}^{(3)}\big({\mathbf k},e;
	{\mathbf k'},o\big) \Big]
	\nonumber\\
	&+\, {\mathrm Re}\Big[{\mathcal A}^{(1) *}\big({\mathbf k},o;
	{\mathbf k'},e\big)\, {\mathcal A}^{(3)}\big({\mathbf k},o;
	{\mathbf k'},e\big) \Big]
	\Big\}\,\frac{ L^d  d^d{\mathbf k}}{(2 \pi)^d} \,
	\frac{L^d d^d{\mathbf k'}}{(2\pi)^d} \;,
\end{align}
where the overall Bose factor has already been
applied; note that it multiplies the squared amplitudes
$|{\mathcal A}^{(2)}|^2$ as well, so each enters with weight $\tfrac12$. Since
${\mathcal A}^{(2)}({\mathbf k},e;{\mathbf k'},e)=
{\mathcal A}^{(2)}({\mathbf k},o;{\mathbf k'},o)$ in the Dirichlet limit, the 
two
square moduli may be merged into a single term
$|{\mathcal A}^{(2)}({\mathbf k},e;{\mathbf k'},e)|^2$.

It is convenient to introduce the loop kernel appearing in
${\mathcal A}^{(2)}$ and the three-vertex object appearing in
${\mathcal A}^{(3)}$,
\begin{align}\label{eq:JPhidef}
	J(k_\shortparallel,k'_\shortparallel) &\equiv
	\int \frac{d^d p_\shortparallel}{(2\pi)^d}\,
	\widetilde{\psi}(k_\shortparallel - p_\shortparallel)\,
	\sqrt{- p_\shortparallel^2}\,
	\widetilde{\psi}(p_\shortparallel + k'_\shortparallel)\;,
	\nonumber\\
	\Phi(k_\shortparallel,k'_\shortparallel) &\equiv
	\tfrac13 (k'^d)^2\,\widetilde{\psi^3}(k_\shortparallel + k'_\shortparallel)
	+ \!\int\! \frac{d^d p_\shortparallel}{(2\pi)^d}
	\widetilde{\psi}(k'_\shortparallel - p_\shortparallel)\,
	p_\shortparallel^2\,
	\widetilde{\psi^2}(p_\shortparallel + k_\shortparallel)
	\nonumber\\
	&\quad
	-2\!\int\! \frac{d^d p_\shortparallel}{(2\pi)^d}
	\frac{d^d q_\shortparallel}{(2\pi)^d}\,
	\widetilde{\psi}(k_\shortparallel - p_\shortparallel)
	\sqrt{- p_\shortparallel^2}\,
	\widetilde{\psi}(p_\shortparallel - q_\shortparallel)
	\sqrt{- q_\shortparallel^2}\,
	\widetilde{\psi}(q_\shortparallel + k'_\shortparallel)\;,
\end{align}
so that, in the Dirichlet limit,
${\mathcal A}^{(2)}({\mathbf k},e;{\mathbf k'},e)= -2i\,{\mathcal N}_k 
{\mathcal N}_{k'}\,k^d k'^d\,
J(k_\shortparallel,k'_\shortparallel)$ and
${\mathcal A}^{(3)}({\mathbf k},e;{\mathbf k'},o)= i\,{\mathcal N}_k {\mathcal 
N}_{k'}\,k^d k'^d\,
\Phi(k_\shortparallel,k'_\shortparallel)$, with ${\mathcal 
N}_k=(L^d|k^0|)^{-1/2}$.
Inserting these into (\ref{eq:dP4master}) and using
${\mathcal A}^{(1)}({\mathbf k},e;{\mathbf k'},o)$ from 
(\ref{eq:eqa1}),
the powers of $L$ cancel against the phase-space measure and one obtains the
explicit fourth-order spectral probability
\begin{align}\label{eq:dP4explicit}
	dP^{(4)}\big({\mathbf k},{\mathbf k'}\big) & =\,
	4\,\frac{(k^d)^2 (k'^d)^2}{|k^0|\,|k'^0|}\,
	\Big[\big|J(k_\shortparallel,k'_\shortparallel)\big|^2
	-\tfrac12\,{\mathrm 
	Re}\!\big(\widetilde{\psi}^{*}(k_\shortparallel+k'_\shortparallel)\,
\Phi(k_\shortparallel,k'_\shortparallel)\big)
\nonumber\\
	& -\tfrac12\,{\mathrm 
	Re}\big(\widetilde{\psi}^{*}(k_\shortparallel+k'_\shortparallel)\,
	\Phi(k'_\shortparallel,k_\shortparallel)\big)\Big]\,
	\frac{d^d{\mathbf k}}{(2\pi)^d}\,\frac{d^d{\mathbf k'}}{(2\pi)^d}\;,
\end{align}
where $k^d=\sqrt{k_\shortparallel^2}$, $k'^d=\sqrt{k'^2_\shortparallel}$ are 
fixed
by the mass-shell conditions, and the two ${\mathrm Re}$ terms are the
$(e,o)$ and $(o,e)$ interferences, which differ only by the exchange
${\mathbf k}\leftrightarrow{\mathbf k'}$ in $\Phi$. Equation
(\ref{eq:dP4explicit}) has a first term which is the modulus squared of the 
one-loop (``box-type'') amplitude
${\mathcal A}^{(2)}$, positive definite; the second collects the
interference of the tree-level amplitude ${\mathcal A}^{(1)}$ with the
three-vertex amplitude ${\mathcal A}^{(3)}$, whose sign makes
the net fourth-order correction to the single-pair spectrum negative on
average (vacuum-persistence depletion, see Sec.~\ref{sec:consistency}).

\subsection{Applications}
We consider here several applications of the general formulas obtained in
the previous subsection.

\subsubsection{Oscillating bump}
\begin{figure}[h]
\centering
\includegraphics[width=0.55\textwidth]{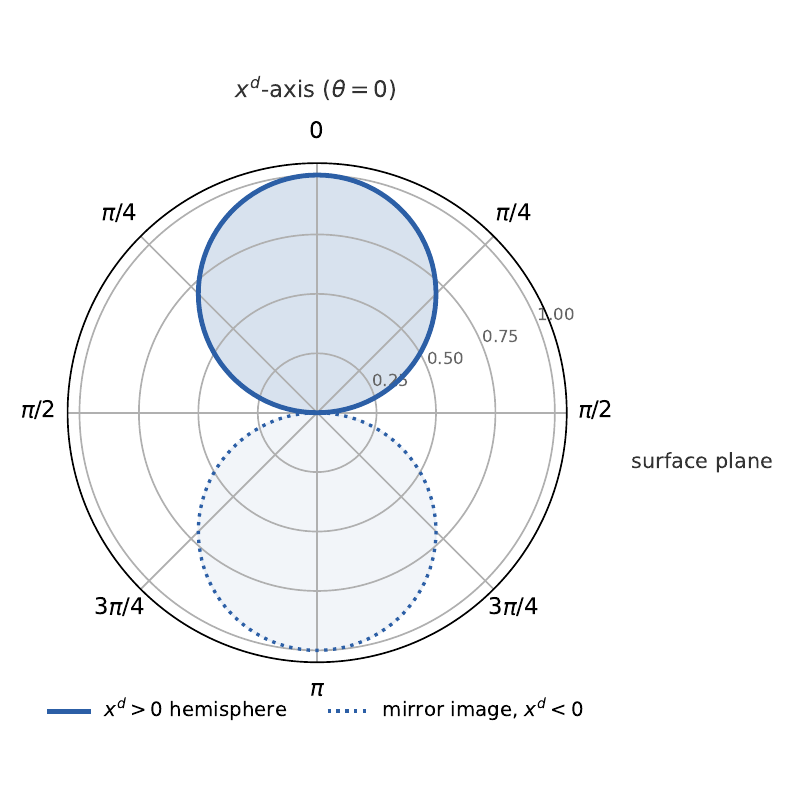}
\caption{Angular distribution of the emitted particles in the localized, 
small-source regime, Eq.~(\ref{eq:angular}): a $\cos\theta$ lobe along 
the normal to the surface. The dotted curve is the mirror image emitted 
into the $x^d<0$ half-space.}\label{fig:angular}
\end{figure}

As a first example, let us evaluate the second-order probability
for a particular case: a small-size deformation, with single frequency $K_s$ 
characterizing its time dependence. By small size, we mean that the 
surface remains planar almost everywhere, except for a region
with small support where it varies. To be more explicit, we assume:
\begin{equation}
\psi(x_\shortparallel) \;=\; \xi({\mathbf x_\shortparallel}) \, \cos(K_s t) 
\;,
\end{equation} 
where $\xi$ is concentrated around a spatial region of size $l$, which we shall 
assume much smaller than the distances from which we want to observe the 
properties of the emitted particles.
Introducing this into (\ref{eq:defp2}), we see that the probability becomes 
proportional to the total evolution time $T$, as it should be for a periodic 
motion. The rate $dW^{(2)} = \frac{dP^{(2)}}{T}$ is: 
\begin{equation}\label{eq:dW21}
dW^{(2)}({\mathbf k}) \,=\,
\frac{d^d{\mathbf k}}{(2\pi)^d} \,|k^d| \, \Theta(K_s - k^0) \,
 \int_U \frac{d^{d-1}{\mathbf k'}}{(2\pi)^{d-1}} \,   
	\big|\widetilde{\xi}({\mathbf {k'}_\shortparallel})\big|^2  \, 
	\sqrt{(K_s - k^0)^2 - ({\mathbf k}_\shortparallel - {\mathbf 
	{k'}_\shortparallel} )^2} \;,
\end{equation}
where
\begin{equation}
U = \{ {\mathbf k'}_\shortparallel : |{\mathbf k}_\shortparallel - {\mathbf 
k'}_\shortparallel | < K_s - k^0 \} .
\end{equation}
Now, we apply the hypothesis regarding the localization of the 
system: at this level, it will mean 
\begin{equation}
({\mathbf k}_\shortparallel - {\mathbf {k'}_\shortparallel} )^2 \,\simeq
\, ({\mathbf {k'}_\shortparallel} )^2 \;,
\end{equation}
an approximation that can be expected to be valid, at least 
for not very large values of $\theta$, the angle between ${\mathbf k}$ and
the $d$ axis (using spherical coordinates in $d$ dimensions).
Recalling that $|{\mathbf k}| = k^0$, we see that:
\begin{equation}\label{eq:dW22}
dW^{(2)}({\mathbf k}) \,\simeq\, \frac{(k^0)^d}{(2\pi)^d} 
\,\cos\theta\, d\Omega\, \Theta(K_s - k^0) \,
\int_U \frac{d^{d-1}{\mathbf k'}}{(2\pi)^{d-1}} \,  
\big|\widetilde{\xi}({\mathbf {k'}_\shortparallel})\big|^2  \, 
\sqrt{(K_s
	 - k^0)^2 - ({\mathbf 
{k'}_\shortparallel} )^2} \;,
\end{equation} 
where $\Omega$ is the solid angle.
The probability of detecting a particle per unit of solid angle
per unit time results from integrating over energies:
\begin{equation}\label{eq:angular}
\frac{dP(\theta)}{d\Omega} \,\simeq\, C\,\cos\theta\,\;,
\end{equation}
with
\begin{equation}
C \, \equiv \, \int_0^{K_s} dk^0 \frac{(k^0)^d}{(2\pi)^d} \int_U 
\frac{d^{d-1}{\mathbf k'}}{(2\pi)^{d-1}} \,  
\big|\widetilde{\xi}({\mathbf {k'}_\shortparallel})\big|^2  \, 
\sqrt{(K_s - k^0)^2 - ({\mathbf {k'}_\shortparallel} )^2} \;.
\end{equation}
The angular pattern $dP/d\Omega \propto \cos\theta$ is the dipole-like 
distribution shown in Fig.~\ref{fig:angular}. The intensity is maximal 
along the normal to the surface ($\theta = 0$) and vanishes tangentially 
($\theta = \pi/2$); the structure is reminiscent of Lambertian emission 
from an aperture and is dictated entirely by the kinematical factor 
$|k^d| = k^0\cos\theta$ in (\ref{eq:dW21}). The pattern is invariant 
under $x^d\to-x^d$, so an identical lobe (dotted in 
Fig.~\ref{fig:angular}) is emitted into the $x^d<0$ half-space.

We conclude the discussion of this example by mentioning that, had
we wished to consider the radiated power, the qualitative features of
the results would have been the same. Indeed, to obtain, for instance,
the radiated energy per unit of solid angle, per unit time, one just
multiplies (\ref{eq:angular}) by $k^0$, which is the energy of the
emitted quanta.

\subsubsection{Oscillating plane}
This is an example which, in its electromagnetic version, has
been considered in~\cite{MaiaNeto1996}, so it is worth considering
the analogous situation here, for the sake of the comparison.
A plane oscillating with an amplitude $b$ and a frequency $K_s$ 
corresponds to $\psi(x_\parallel) = b \cos(K_s t)$,

Inserting this into the general expression for the probability, and multiplying
by the energy of the emitted quanta, we get the power per unit of
solid angle, energy, and per area of the oscillating plane:
\begin{equation}
\frac{1}{A_\shortparallel}\, \frac{dW^{(2)}}{dk^0 \,d\Omega} 
\,=\, \frac{b^2 (k^0)^5}{(2 \pi)^3} \, 
\big(\frac{k^0}{K_s}\big)^4 \, \cos^2\theta  \, 
\sqrt{ 1 - 2 \frac{k^0}{K_s} + \big(\frac{k^0}{K_s}\big)^2 
\cos^2\theta} \;.
\end{equation}
The $\cos^2\theta$ profile and the kinematic square-root edge reproduce, for 
the scalar field, the structure found for the electromagnetic field 
in~\cite{MaiaNeto1996}: our Dirichlet result is the analog of their 
transverse-electric (Dirichlet) polarization, the twin quanta being emitted 
with opposite parallel momenta and energies adding up to the drive frequency 
$K_s$, exactly as in their twin-photon spectra (their transverse-magnetic, 
i.e.\ Neumann, channel would correspond to the complementary boundary 
condition).

\subsubsection{Second-harmonic and complementary first-harmonic emissions at  
fourth order}
The single-frequency drive makes the harmonic content of each amplitude 
explicit: every insertion of $\widetilde{\psi}$ delivers an energy $\pm K_s$ at 
vanishing transverse momentum, so ${\mathcal A}^{(1)}$ feeds pairs with 
$k^0+k'^0=K_s$, while ${\mathcal A}^{(2)}$, which carries two insertions, feeds 
pairs with $k^0+k'^0=2K_s$. Inspecting the fourth-order probability 
(\ref{eq:dP4master}) channel by channel, the box term 
$|{\mathcal A}^{(2)}|^2$ is supported \emph{entirely} on the second harmonic 
$k^0+k'^0=2K_s$, a process kinematically forbidden at second order,
whereas the interference $2\,{\rm Re}\,[{\mathcal A}^{(1)*}{\mathcal A}^{(3)}]$ 
stays at the first harmonic $K_s$ and merely provides an $O(b^4)$ correction to 
the leading rate, negative on average (the vacuum-persistence depletion 
discussed in Sec.~\ref{sec:consistency}). The genuinely new fourth-order 
feature is therefore the opening of the $2K_s$ line.

The second-harmonic channel is self-contained, as it receives no interference, 
and can be evaluated in closed form. Inserting 
$\widetilde{\psi}(p_\shortparallel)$ into the loop kernel 
$J$ of (\ref{eq:JPhidef}) forces the transverse momenta to be opposite, 
$\mathbf{k}+\mathbf{k}'=0$, and the energies to share the second harmonic, 
$k^0+k'^0=2K_s$; the loop momentum left over is 
$p_\shortparallel^\star=(k^0-K_s,\mathbf{k})$, whose weight in 
$|{\mathcal A}^{(2)}|^2$ is 
$|\!-\!p_\shortparallel^{\star 2}|=|\mathbf{k}^2-(k^0-K_s)^2|$. Writing 
$u\equiv k^0/K_s$ and $r\equiv|\mathbf{k}|/K_s$, the total second-harmonic rate 
per unit transverse area of the plane is
\begin{equation}\label{eq:P4_2h}
	\frac{P^{(4)}_{2K_s}}{A_\shortparallel\,T}\,=\,
	\frac{4\pi^4\,\Omega_{d-2}}{(2\pi)^{d+4}}\,{\mathcal I}_d\;b^4 K_s^{d+4}\;,
\end{equation}
with $\Omega_{d-2}=2\pi^{(d-1)/2}/\Gamma(\tfrac{d-1}{2})$ the area of the unit 
$(d{-}2)$-sphere and
\begin{equation}\label{eq:Id2h}
	{\mathcal I}_d\,=\,\int_0^2 du\int_0^{\min(u,2-u)}\!\!dr\;
	r^{d-2}\,\sqrt{u^2-r^2}\,\sqrt{(2-u)^2-r^2}\;\big|\,r^2-(u-1)^2\,\big|\;.
\end{equation}
The $d=1$ case, lacking transverse momentum, is handled directly. The angular 
integral is rational in low dimensions 
(${\mathcal I}_2=\tfrac19$, ${\mathcal I}_4=\tfrac1{45}$, and the $d=1$ 
reduction yields $\tfrac{4}{15}$), so that
\begin{equation}\label{eq:P4_2h_cases}
	\frac{P^{(4)}_{2K_s}}{A_\shortparallel T}\,=\,
	\begin{cases}
		\;\dfrac{b^4K_s^5}{30\pi}\,, & d=1\,,\\[2.0ex]
		\;\dfrac{b^4K_s^6}{72\pi^2}\,, & d=2\,,\\[2.0ex]
		\;\dfrac{{\mathcal 
		I}_3}{16\pi^2}\,b^4K_s^7\,\simeq\,2.54\times10^{-4}\,b^4K_s^7\,, 
		& d=3\,,\\[2.0ex]
		\;\dfrac{b^4K_s^8}{720\pi^3}\,, & d=4\,,
	\end{cases}
\end{equation}
with ${\mathcal I}_3=0.040053\ldots$ (no elementary closed form). The new 
channel is suppressed relative to the leading emission by the square of the 
expansion parameter $bK_s\sim v_{\rm max}$, the maximal velocity of the 
surface; 
in $d=1$ the comparison with the leading rate (\ref{eq:Ptot_finite_lambda}) is 
exact,
\begin{equation}\label{eq:ratio_2h}
	\frac{P^{(4)}_{2K_s}}{P^{(2)}_{K_s}}\,=\,\frac{2}{5}\,(bK_s)^2\,.
\end{equation}
Physically, the plane driven at $K_s$ radiates twin quanta that share the 
energy $2K_s$ with opposite transverse momenta, the second-harmonic 
counterpart of the first-harmonic twin photons of~\cite{MaiaNeto1996}, and 
its $b^4$ scaling identifies it as a genuinely nonlinear imprint of the surface 
dynamics.

The complementary, first-harmonic piece of the fourth-order probability is the 
interference $2\,{\rm Re}[{\mathcal A}^{(1)*}{\mathcal A}^{(3)}]$, which by the 
same harmonic counting remains at $k^0+k'^0=K_s$ and thus corrects the 
\emph{leading} rate at $O(b^4)$; it is the explicit, oscillating-plane face of 
the vacuum-persistence depletion of Sect.~\ref{sec:consistency}. In $d=1$ it 
can 
be evaluated in closed form: every momentum is an energy, $k^d=k^0$, the loop 
factors enter as real pairs, 
$\sqrt{-p_\shortparallel^2}\,\sqrt{-q_\shortparallel^2}=-|p^0||q^0|$, and the 
three contributions to $\Phi$ in (\ref{eq:JPhidef}) --- the 
$\widetilde{\psi^3}$ term, the $\widetilde{\psi}\,\widetilde{\psi^2}$ term and 
the triple-$\widetilde{\psi}$ convolution --- collapse onto the first harmonic. 
With $u\equiv k^0/K_s\in[0,1]$,
\begin{equation}\label{eq:Phi1h}
	\Phi\big|_{K_s}\,=\,\frac{\pi b^3}{4}\,\big(6u^2-8u+9\big)\,K_s^2\;
	\delta(k^0+k'^0-K_s)\;.
\end{equation}
Inserting this into (\ref{eq:dP4explicit}) and integrating over the energy 
sharing gives the $O(b^4)$ correction to the first-harmonic emission rate,
\begin{equation}\label{eq:gamma4_1h}
	\frac{P^{(4)}_{K_s}}{T}\,=\,-\,\frac{17}{120\pi}\,b^4K_s^5
	\,=\,-\,\frac{17}{10}\,(bK_s)^2\,\frac{P^{(2)}}{T}\,,
\end{equation}
negative, as anticipated. The two fourth-order effects are therefore explicit 
in $d=1$: the first-harmonic line is depleted by a relative 
$O\big((bK_s)^2\big)$, Eq.~(\ref{eq:gamma4_1h}), while a second-harmonic line 
opens with the positive rate (\ref{eq:P4_2h_cases}). Both scale as $b^4K_s^5$ 
and are controlled by the same expansion parameter $bK_s\sim v_{\rm max}$; 
their 
sum, the net connected fourth-order rate, is $-\tfrac{13}{120\pi}\,b^4K_s^5$.

\subsubsection{One spatial dimension ($d=1$)} \label{subsec:d1}
This is a useful example in its own right, since the lack of angular dependence allows
for the evaluation of rather general expressions, even for a finite 
$\lambda$.  
Here, we see that $k^0$ and $k^d = k^1$ are both positive and identical. Then
\begin{equation}
dP^{(2)}(k^1) \,=\,dP^{(2)}(k^0) \,=\, 4 \frac{dk^1}{2\pi}
\int_0^\infty \,\frac{d{k'}^1}{2\pi}\, 
\big|\widetilde{\psi}(k^1 + {k'}^1)\big|^2  \, k^1 {k'}^1\;.
\end{equation}
Thus, we may write the spectral probability density, as follows:
\begin{align}
\frac{dP^{(2)}(\omega)}{d\omega}  & =\,\frac{1}{\pi^2} \,
\int_0^\infty d\nu \, 
\big|\widetilde{\psi}(\omega + \nu)\big|^2  \, \omega \, \nu \nonumber\\
 & =\,\frac{1}{\pi^2} \,\omega \,  \int_\omega^\infty d\nu \, 
\big|\widetilde{\psi}(\nu)\big|^2  \, (\nu - \omega) \;.
\end{align}
In the particular case of harmonic motion, 
\begin{equation}
	\psi(t) \,=\, b \, \cos(K_s t) 
\end{equation}
the probability density per unit time is:
\begin{equation}
	\gamma(\omega) \, \equiv 
	\,\Big[\frac{1}{T}\frac{dP^{(2)}(\omega)}{d\omega}\Big]_{T \to \infty} 
	\,=\,\frac{b^2}{2\pi}\, \Theta(K_s - \omega) \; \omega \, 
	(K_s - \omega) \;. 
\end{equation}

Finally, for any deformation, we can always consider
the total probability, obtained by integrating over the possible 
values of $\omega$,
\begin{align}\label{eq:Ptot_d1_Dir}
\int d\omega \frac{dP^{(2)}(\omega)}{d\omega}  &=\,
\frac{1}{\pi^2}  \,  \int_0^\infty d\nu \, 
\big|\widetilde{\psi}(\nu)\big|^2  \,\int_0^\infty d\omega \, 
\omega\, \Theta(\nu - \omega) \, (\nu - \omega) \nonumber\\
	& =\,\frac{1}{6\pi^2}  \,  \int_0^\infty d\nu \, 
\big|\widetilde{\psi}(\nu)\big|^2  \,\nu^3 
=\,\frac{1}{12\pi^2}  \,  \int_{-\infty}^\infty d\nu \, 
\big|\widetilde{\psi}(\nu)\big|^2  \,|\nu|^3
\;.
\end{align}

\begin{figure}[h]
\centering
\includegraphics[width=0.78\textwidth]{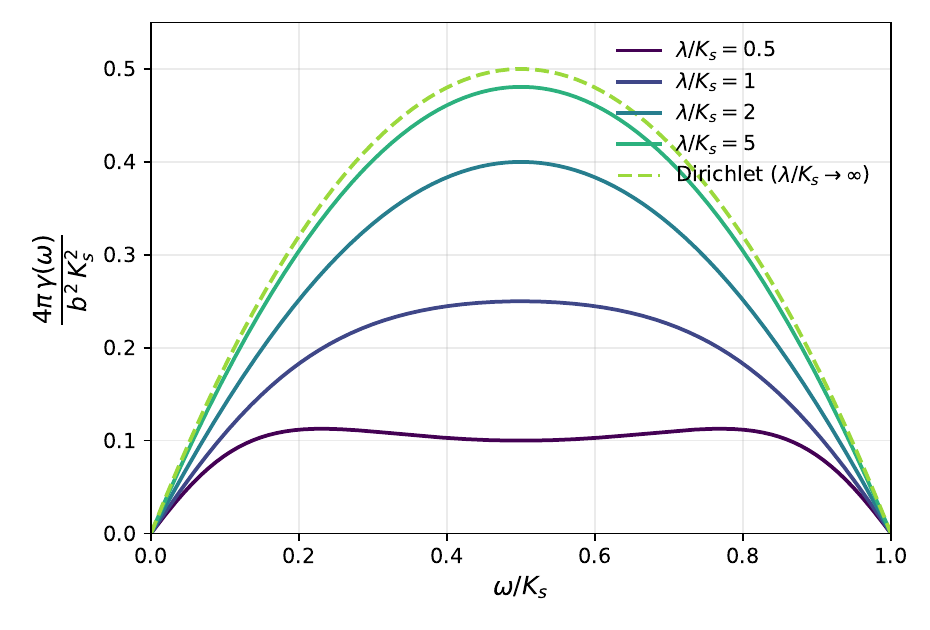}
\caption{Pair-creation rate $\gamma(\omega)$ of 
Eq.~(\ref{eq:gamma_finite_lambda}) (in units of $b^2 K_s^2/(4\pi)$) 
as a function of $\omega/K_s$, for several values of 
$\lambda/K_s$. The Dirichlet limit (dashed) is the parabola 
$2(\omega/K_s)(1-\omega/K_s)$. For $\lambda\lesssim K_s$ the 
spectrum departs from the Dirichlet shape and develops a mild bimodal 
structure, the Lorentzian factor 
$[1+(2k_\shortparallel/\lambda)^2]^{-1}$ favoring asymmetric 
partitions of the energy $K_s$.}\label{fig:gamma_finite_lambda}
\end{figure}

For a finite $\lambda$, and the same oscillatory configuration we had
before, the probability density per unit time is:
\begin{equation}\label{eq:gamma_finite_lambda}
\gamma(\omega) \,=\,\frac{b^2}{4\pi}\, 
\Theta(K_s - \omega) \; \omega \, (K_s - \omega) \,
\Big\{\big[ 1 + (\frac{2 \omega}{\lambda} )^2\big]^{-1} + \big[ 1 + (\frac{2 
(K_s - \omega)}{\lambda} )^2\big]^{-1} \Big\}\;. 
\end{equation}
The shape of $\gamma(\omega)$ is shown in Fig.~\ref{fig:gamma_finite_lambda}
for several values of the dimensionless ratio $\lambda/K_s$.

Integrating $\gamma(\omega)$ over $\omega\in[0,K_s]$ yields the total
probability per unit time, which admits the closed form
\begin{equation}\label{eq:Ptot_finite_lambda}
	\frac{P^{(2)}}{T} \,=\, \frac{b^{2}K_s^{3}}{4\pi}\,
	\bigg[\frac{r^{2}}{4}\ln\!\Big(1+\frac{4}{r^{2}}\Big)
	\,-\,\frac{r^{2}}{2}\,+\,\frac{r^{3}}{4}\arctan\!\frac{2}{r}\bigg]\;,
	\qquad r\equiv\frac{\lambda}{K_s}\;.
\end{equation}
This expression interpolates between the weak-coupling regime, 
$\,P^{(2)}/T \simeq (b^{2}K_s^{3}/8\pi)\,r^{2}\ln(2/r)\,$ for 
$\lambda\ll K_s$, and the Dirichlet limit, 
$\,P^{(2)}/T \to b^{2}K_s^{3}/(12\pi)$ for $\lambda\gg K_s$, with 
$O(r^{-2})$ corrections.

%%%%%%%%%%%%%%%%%%%%%%%%%%%%%%%%%%%%%%%%%%%%%%%%%%%%%%%%%%%%%%%%%%%%%%%%%%%
%%%%%%%%%%%%%%%%%%%%%%%%%%%%%%%%%%%%%%%%%%%%%%%%%%%%%%%%%%%%%%%%%%%%%%%%%%%
%%%%%%%%%%%%%%%%%%%%%%%%%%%%%%%%%%%%%%%%%%%%%%%%%%%%%%%%%%%%%%%%%%%%%%%%%%%
%%%%%%%%%%%%%%%%%%%%%%%%%%% Consistency %%%%%%%%%%%%%%%%%%%%%%%%%%%%%%%%%%%
%%%%%%%%%%%%%%%%%%%%%%%%%%%%%%%%%%%%%%%%%%%%%%%%%%%%%%%%%%%%%%%%%%%%%%%%%%%
%%%%%%%%%%%%%%%%%%%%%%%%%%%%%%%%%%%%%%%%%%%%%%%%%%%%%%%%%%%%%%%%%%%%%%%%%%%
%%%%%%%%%%%%%%%%%%%%%%%%%%%%%%%%%%%%%%%%%%%%%%%%%%%%%%%%%%%%%%%%%%%%%%%%%%%
\section{Consistency with previous results}\label{sec:consistency}
In order to compare our results for the imaginary part of the effective action
$\Gamma$ with the ones obtained in~\cite{FG2024}, we shall denote by $\Gamma_{[\mathrm{FG}]}$ the
effective action used as reference in this comparison. We first carefully
consider the relation between the imaginary part of the effective action and the
exclusive probability of emission of a pair. In particular, we shall see that
$P^{(2)}=2\,{\rm Im}\,\Gamma^{(2)}$ is modified in an instructive way, since a
\emph{four-particle} (two-pair) channel opens at precisely the same order in
$\psi$.

We shall start from the observation that, for an interaction quadratic in
$\varphi$, the reduction 
formulas~\cite{Lehmann:1954rq} imply that the only connected amplitudes 
involve two particles. The full amplitude is the exponential of the 
connected parts~\cite{Itzykson:1980rh}; the resulting Gaussian (Bogoliubov, 
squeezed-vacuum) structure of the in-out $S$-matrix is standard for theories 
quadratic in the field~\cite{BirrellDavies1982}. We use that 
property to write the action of the $S$-matrix on the vacuum in the 
following way:
\begin{equation}\label{eq:squeezed}
S\,|0\rangle \,=\, {\mathcal V}\,
\exp\!\Big(\tfrac12\sum_{i,j}{\mathcal A}(i ; j)\, a^\dagger_i 
a^\dagger_j\Big)|0\rangle\;,
\qquad {\mathcal V}\equiv\langle 0|S|0\rangle\;,
\end{equation}
where we have used a shorthand notation for the exact two-particle amplitudes, 
and the creation operators; i.e., the latter create a quantum in the mode $i 
\equiv({\mathbf k},\alpha)$. The kernel ${\mathcal A}(i;j)$ is nothing other
than the connected pair amplitude of Sec.~\ref{sec:results} (it is symmetric 
under interchange of its indices).

Projecting (\ref{eq:squeezed}) on a two-particle state:
\begin{equation}\label{eq:Omega_vs_A}
	{\mathcal A}(i;j)\,=\,\frac{\langle i , j|S|0\rangle}{\langle 0|S|0\rangle}
	\,=\,{\mathcal A}^{(1)}(i;j)+{\mathcal A}^{(2)}(i;j)+{\mathcal 
	A}^{(3)}(i;j)+\ldots\;,
\end{equation}	
where we introduced the perturbative expansion for each coefficient. This is the very amplitude defined in (\ref{eq:Adef}): the division by ${\mathcal V}=\langle 0|S|0\rangle$ in (\ref{eq:Omega_vs_A}) is exactly the removal of the disconnected vacuum subdiagrams anticipated there, so the kernel in (\ref{eq:squeezed}) is the one whose perturbative coefficients ${\mathcal A}^{(1)},{\mathcal A}^{(2)},{\mathcal A}^{(3)}$ were computed in Sec.~\ref{sec:results}.

Projecting (\ref{eq:squeezed}) on four-particle states yields,
\begin{equation}\label{eq:fourpt}
	\langle i,j,k,l|S|0\rangle\,=\,{\mathcal V}\,\big[\,
	{\mathcal A}(i;j){\mathcal A}(k;l)+{\mathcal A}(i;k){\mathcal A}(j;l)
	+{\mathcal A}(i;l){\mathcal A}(j;k)\,\big]\;,
\end{equation}
i.e., the four-particle (or two-pair) amplitude is the symmetrized product of 
pair amplitudes (the three terms are the three ways of grouping four identical 
bosons into two pairs).

So far, we have not used the property that $S$ is unitary. This fixes the 
modulus of ${\mathcal V}$
\begin{equation}\label{eq:v}
|{\mathcal V}|^2 \, \big(1 + {\mathcal P} + \tfrac12 {\mathcal P}^2 + {\mathcal 
Q} + 
\ldots \big) = 1
\end{equation} 
where:
\begin{align}
{\mathcal P} &=\, \tfrac12 \sum_{i,j}  |{\mathcal A}(i;j)|^2 \nonumber\\
{\mathcal Q} &=\, \tfrac14 \sum_{i,j,k,l} {\mathcal A}^*(i;j) {\mathcal A}(j;k)
 {\mathcal A}^*(k;l) {\mathcal A}(l;i) \;,
\end{align}
and we kept terms with up to four powers of the 
amplitudes. We have abbreviated $\sum_i\equiv\sum_{\alpha=o,e}\int L^d 
d^d{\mathbf k}/(2\pi)^d$, so that every 
sum over indices above involves a phase-space integral plus a sum over the $e$ 
and $o$ indices.

With the above notation, we see that the {\em exact\/} probability of 
detecting a 
single pair, $p_1$, and two pairs, $p_2$, are then written in a rather compact 
fashion:
\begin{equation}
p_1  \,=\,  |{\mathcal V}|^2 {\mathcal P} \;\;,\;\;\;
p_2  \,=\,  |{\mathcal V}|^2 \big( \tfrac12 {\mathcal P}^2 
+ {\mathcal Q} \big) \;,
\end{equation}
which, using (\ref{eq:v}), leads to:
\begin{equation}
	p_1  \,=\, {\mathcal P} - {\mathcal P}^2 + \ldots\;\;,\;\;\;
	p_2  \,=\,  \tfrac12 {\mathcal P}^2 
	+ {\mathcal Q} + \ldots \;,
\end{equation}
where we have kept terms with up to four amplitudes. We remark that except for
the latter, no perturbative expansion for the amplitudes themselves has been 
used yet. Expanding the amplitudes perturbatively,
\begin{align}
p_1^{(2)}  & =\, {\mathcal P}^{(2)} \;\;,\;\;\;
p_1^{(4)} \,=\, {\mathcal P}^{(4)} - ({\mathcal P}^{(2)})^2 \;,\nonumber\\
p_2^{(4)}  &=\, \tfrac12 ({\mathcal P}^{(2)})^2  + {\mathcal Q}^{(4)} \;.  	
\end{align}
Finally, to check against the imaginary part of the effective 
action $\Gamma$, we note that	
	\begin{equation}\label{eq:ImGamma_trace}
	2\,{\rm Im}\,\Gamma \,=\, -\ln|{\mathcal V}|^2
	\,=\, {\mathcal P} \,+\, {\mathcal Q} \,+\,\ldots\;.
\end{equation}
Introducing the expansion of $\Gamma$, one obtains:
\begin{equation}
2\,{\rm Im}\Gamma^{(2)} \,=\, {\mathcal P}^{(2)}
\;\;,\;\;
2\,{\rm Im}\Gamma^{(4)} \,=\, {\mathcal P}^{(4)} + {\mathcal Q}^{(4)}
\end{equation}
It follows that
\begin{equation}
{\mathcal P}^{(2)} = 2\,{\rm Im}\,\Gamma^{(2)}=\tfrac12\sum_{i,j}|{\mathcal 
	A}^{(1)}(i;j)|^2\,,
\end{equation}
which we verify explicitly below.

On the other hand, for the total fourth-order decay probability of the vacuum,
we need to add the corresponding probabilities at that order. 
We find:
\begin{equation}\label{eq:P4decay}
p_1^{(4)}+p_2^{(4)}
		\,=\, 2\,{\rm Im}\,\Gamma^{(4)} \,-\, 2\big({\rm 
		Im}\,\Gamma^{(2)}\big)^2
\end{equation}
which is exactly the $O(\psi^4)$ term of $1-e^{-2\,{\rm Im}\,\Gamma}$. 
The nonlinear subtraction $-2({\rm Im}\,\Gamma^{(2)})^2$ embodies the 
familiar fact that $2\,{\rm Im}\,\Gamma$ is not the probability of producing a 
pair but the (mean) inclusive yield: the two coincide only to leading order and 
differ once the multi-pair channels are populated, the vacuum being 
correspondingly depleted~\cite{Schwinger1951,Nikishov1970,CohenMcGady2008}.

The second- and fourth-order terms in the expansion of ${\mathcal P}$ are obtained by integrating the differentials we have already calculated; 
explicitly, ${\mathcal P}^{(2)}=P^{(2)}\equiv\int dP^{(2)}$ and 
${\mathcal P}^{(4)}=P^{(4)}\equiv\int dP^{(4)}$ are the integrated spectral 
probabilities (\ref{eq:defp2}), (\ref{eq:dP4explicit}) of 
Sec.~\ref{sec:results}, the calligraphic symbol merely emphasizing their 
origin in the unitarity sum (\ref{eq:v}).
On the other hand, 
\begin{equation}\label{eq:Q4}
	{\mathcal Q}^{(4)}\,=\,\tfrac14\sum_{i,j,k,l}
	{\mathcal A}^{(1)*}(i;j)\,{\mathcal A}^{(1)}(j;k)\,
	{\mathcal A}^{(1)*}(k;l)\,{\mathcal A}^{(1)}(l;i)\;.
\end{equation}

The comparison at second order can then be summarized in the relation
\begin{equation}\label{eq:overall}
	P^{(2)} \,=\, 2\, \mathrm{Im}\big[\Gamma^{(2)}_{[\textrm{FG}]}\big]\;,
\end{equation}
the leading-order form of the standard relation $P=2\,{\rm Im}\,\Gamma$.

For any dimension $d$, integrating (\ref{eq:dPgenD}) over the two outgoing momenta we get
\begin{equation}\label{eq:Ptot1}
	P^{(2)} \,=\, \int dP^{(2)}\big(k_\shortparallel,{k'}_\shortparallel\big)
	 \,=\, 4 \,
	\int \frac{d^dk_\shortparallel}{(2\pi)^d}\,\theta(k^0)\,
	\sqrt{k_\shortparallel^2} \, \int 
	\frac{d^dk'_\shortparallel}{(2\pi)^d}\,
	\big|\widetilde{\psi}(k_\shortparallel + 
	{k'}_\shortparallel)\big|^2 \,\theta({k'}^0)\, \sqrt{{k'}_\shortparallel^2} 
 \;,
\end{equation}
where the on-shell positivity of $k_\shortparallel^2$ and 
${k'}_\shortparallel^2$ is implicit in the requirement that the square roots
be real.

The double integral may be reduced to a \emph{single} integral over the 
parallel momentum $q\equiv k_\shortparallel+{k'}_\shortparallel$ carried 
by~$\psi$. Performing the shift $k'_\shortparallel \to q-k_\shortparallel$ at
fixed $k_\shortparallel$,
\begin{equation}\label{eq:Pas-singleQ}
	P^{(2)}\,=\,4 \int \frac{d^d q}{(2\pi)^d}\,
	\big|\widetilde{\psi}(q)\big|^2 \,F_d(q)\;,
\end{equation}
where the kernel
\begin{equation}\label{eq:Fdef}
	F_d(q) \,\equiv\, \int \frac{d^d k_\shortparallel}{(2\pi)^d}\,
	\Theta(k^0)\sqrt{k_\shortparallel^2}\;
	\Theta(q^0-k^0)\sqrt{(q-k_\shortparallel)^2}
\end{equation}
is a Lorentz-invariant function of $q$, supported on the forward light cone 
of the $d$-dimensional parallel space-time. By Lorentz invariance and 
dimensional analysis,
\begin{equation}\label{eq:Fclosed}
	F_d(q)\,=\,c_d\,(q^2)^{(d+2)/2}\,\Theta(q^0)\,\Theta(q^2)\;,
\end{equation}
with a dimensionless coefficient $c_d$ which we compute now.

In the rest frame of $q^\mu = (Q,\mathbf{0})$, $Q=\sqrt{q^2}$, the standard 
spectral representation
\begin{equation}
	\Theta(k^0)\theta(k^2)\sqrt{k^2}\,=\,\int_0^\infty 2 m^2\,dm\;
	\Theta(k^0)\,\delta(k^2-m^2)
\end{equation}
reduces (\ref{eq:Fdef}) to a weighted two-particle phase-space integral.
After scaling $m=Q x$, $m'=Q y$ with $0\le x,y$, $x+y\le 1$, one obtains
\begin{equation}\label{eq:Fass}
	F_d(q)\,=\,\frac{(q^2)^{(d+2)/2}}{2^{2d-4}\,\pi^{(d+1)/2}\,
	\Gamma\!\big((d-1)/2\big)}\,I_d\,\Theta(q^0)\Theta(q^2)\;,
\end{equation}
\begin{equation}\label{eq:Idef}
	I_d \,\equiv\, \iint_{\substack{x,y\ge 0 \\ x+y\le 1}} 
	dx\,dy\;x^2 y^2\,
	\Big[\big(1-(x+y)^2\big)\big(1-(x-y)^2\big)\Big]^{(d-3)/2}\;,
\end{equation}
where the last bracket is the K\"all\'en triangle function for the on-shell
two-particle decomposition. Therefore
\begin{equation}\label{eq:cdformula}
	c_d \,=\,\frac{I_d}{2^{2d-4}\,\pi^{(d+1)/2}\,
	\Gamma\!\big((d-1)/2\big)}\;.
\end{equation}
The integral $I_d$ is rational for odd~$d$ (where $(d-3)/2$ is a 
non-negative integer) and a rational multiple of $\pi^2$ for even~$d$. The 
case $d=1$ is degenerate in the rest-frame derivation but follows from the 
direct one-dimensional calculation already performed in Sec.~\ref{subsec:d1}.
Combining (\ref{eq:Pas-singleQ}) and (\ref{eq:Fass}), the second-order 
total decay probability in the Dirichlet limit reads, in arbitrary 
dimension, 
\begin{equation}\label{eq:Pdfinal}
	\;P^{(2)}\,=\,\mathcal{C}_d\,\int d^d q\,
	\Theta(q^0)\,\Theta(q^2)\,\big|\widetilde{\psi}(q)\big|^2\,
	(q^2)^{(d+2)/2}\;,\quad 
	\mathcal{C}_d \,=\,
	\frac{8^{2-d} \,\pi^{-(3d+1)/2}I_d}{\Gamma\!\big((d-1)/2\big)}\;.
\end{equation}
For $d=1$, Eq.~(\ref{eq:Pdfinal}) reproduces the result quoted in 
Sec.~\ref{subsec:d1}, in full agreement with Eq.~(\ref{eq:Ptot_d1_Dir}).

We can now make the comparison advertised in (\ref{eq:overall}) fully 
explicit. The second-order imaginary part of the effective action obtained 
in~\cite{FG2024} can be written, for arbitrary $d$, as
\begin{equation}\label{eq:ImGFG}
	\mathrm{Im}\big[\Gamma^{(2)}_{[\textrm{FG}]}\big] 
	\,=\, \frac{\eta_d}{2}\,\int \frac{d^d k_\shortparallel}{(2\pi)^d}\,
	\Theta\big(k_\shortparallel^2\big)\,
	\big(k_\shortparallel^2\big)^{(d+2)/2}\,
	\big|\widetilde{\psi}(k_\shortparallel)\big|^2\;,
\end{equation}
where $k_\shortparallel^2 \equiv (k^0)^2 - {\mathbf k}_\shortparallel^2$. 
The Heaviside function $\Theta(k_\shortparallel^2)$ encompasses both 
branches of the (parallel) light cone, and the dimensionless coefficient 
$\eta_d$ is given, in odd and even spatial dimensions, by [cf.\ Eqs.~(41) 
and~(44) of~\cite{FG2024}]
\begin{equation}\label{eq:etadef}
	\eta_{2q+1}\,=\,\frac{(q!)^2}{(2\pi)^{q+1}\,(2q+1)!\,(2q+3)!!}\;,
	\qquad
	\eta_{2q}\,=\,\frac{\pi}{(4\pi)^q}\,
	\frac{\big[\Gamma(q+\tfrac{1}{2})\big]^2}
	{(q+1)!\,(2q)!\,\big[\Gamma(-\tfrac{1}{2})\big]^2}\;.
\end{equation}
Since $\psi$ is real, $|\widetilde\psi(-k_\shortparallel)|^2 = 
|\widetilde\psi(k_\shortparallel)|^2$, and the integrand of (\ref{eq:ImGFG}) 
is symmetric under $k^0\to -k^0$. Hence the integral over the full 
light-cone support breaks into two equal contributions, one over each 
branch:
\begin{equation}\label{eq:halfsymmetry}
	\int d^d k_\shortparallel\,\Theta\big(k_\shortparallel^2\big)\,
	\big(k_\shortparallel^2\big)^{(d+2)/2}\,|\widetilde{\psi}|^2 \,=\, 
	2\,\int d^d q\,\Theta(q^0)\,\Theta(q^2)\,(q^2)^{(d+2)/2}\,
	|\widetilde{\psi}(q)|^2\;,
\end{equation}
so that, combining (\ref{eq:overall}), (\ref{eq:ImGFG}) and 
(\ref{eq:halfsymmetry}),
\begin{equation}\label{eq:fourImG}
	2\,\mathrm{Im}\big[\Gamma^{(2)}_{[\textrm{FG}]}\big]
	\,=\,\frac{2\,\eta_d}{(2\pi)^d}\,\int d^d q\,
	\Theta(q^0)\,\Theta(q^2)\,(q^2)^{(d+2)/2}\,
	\big|\widetilde{\psi}(q)\big|^2\;.
\end{equation}
Comparing with our result (\ref{eq:Pdfinal}), the consistency 
$P^{(2)}=2\,\mathrm{Im}\big[\Gamma^{(2)}_{[\textrm{FG}]}\big]$ amounts to 
the single identity
\begin{equation}\label{eq:Cdvseta}
	\;\mathcal{C}_d \,=\, \frac{2\,\eta_d}{(2\pi)^d}\;,
\end{equation}
which we verified for general $d$. Thus, the exclusive 
amplitudes computed here reproduce, upon summing over final states, 
exactly twice the imaginary part of the effective action 
of~\cite{FG2024}, in agreement with the leading order of the standard 
relation $P=2\,\mathrm{Im}\,\Gamma$, i.e.\ 
$P^{(2)}=2\,\mathrm{Im}\,\Gamma^{(2)}$.

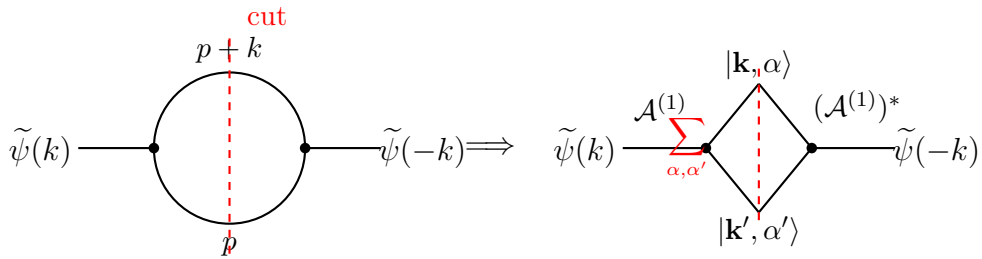
\begin{figure}[h]
\centering
\begin{tikzpicture}[scale=1.0,
	vert/.style={circle,fill,inner sep=1.4pt},
	>=stealth]
  % bubble
  \draw[thick] (-2.0,0) -- (-1.0,0);
  \draw[thick] ( 1.0,0) -- ( 2.0,0);
  \draw[thick] (0,0) circle (1);
  \node[vert] at (-1,0) {};
  \node[vert] at  (1,0) {};
  \node at (-2.5,0) {$\widetilde{\psi}(k)$};
  \node at  ( 2.55,0) {$\widetilde{\psi}(-k)$};
  \node at (0, 1.30) {\small $p+k$};
  \node at (0,-1.30) {\small $p$};
  % cut
  \draw[red,thick,dashed] (0,1.45) -- (0,-1.45);
  \node[red] at (0.50,1.75) {\small cut};
  % equality
  \node at (3.45,0) {$\;\;\Longrightarrow\;\;$};
  % right-hand side: two amplitudes glued at a vertical cut
  \begin{scope}[xshift=7.0cm]
    \draw[thick] (-1.8,0) -- (-0.7,0);
    \draw[thick] (-0.7,0) -- ( 0, 0.85);
    \draw[thick] (-0.7,0) -- ( 0,-0.85);
    \draw[thick] ( 0.7,0) -- ( 0, 0.85);
    \draw[thick] ( 0.7,0) -- ( 0,-0.85);
    \draw[thick] ( 0.7,0) -- ( 1.8,0);
    \node[vert] at (-0.7,0) {};
    \node[vert] at  ( 0.7,0) {};
    \node at (-2.3,0) {$\widetilde{\psi}(k)$};
    \node at  ( 2.35,0) {$\widetilde{\psi}(-k)$};
    \node at ( 0, 1.10) {\small $|{\mathbf k},\alpha\rangle$};
    \node at ( 0,-1.10) {\small $|{\mathbf k}',\alpha'\rangle$};
    \draw[red,thick,dashed] (0,1.05) -- (0,-1.05);
    \node[red] at (-0.95,0) {\footnotesize $\displaystyle\sum_{\alpha,\alpha'}$};
    \node at (-1.3,0.55) {\small ${\mathcal A}^{(1)}$};
    \node at ( 1.3,0.55) {\small $({\mathcal A}^{(1)})^{*}$};
  \end{scope}
\end{tikzpicture}
\caption{Cutting rule for the second-order vacuum-decay probability. The 
discontinuity of the bubble diagram for 
$\Gamma^{(2)}_{[\mathrm{FG}]}$ (left) across its two internal lines 
equals the on-shell two-particle phase-space integral of 
$|{\mathcal A}^{(1)}|^2$ (right), with the sum running over the parities 
$(\alpha,\alpha')\in\{(e,o),(o,e)\}$ of the emitted quanta.}\label{fig:cut}
\end{figure}

This identity is the dynamical-Casimir realization of Cutkosky's cutting 
rule: the discontinuity of the bubble diagram for 
$\Gamma^{(2)}_{[\mathrm{FG}]}$ across the two-particle threshold equals 
the on-shell phase-space integral of $|{\mathcal A}^{(1)}|^2$. 
Pictorially (Fig.~\ref{fig:cut}), cutting the loop along its two internal 
lines yields, on each side of the cut, a copy of the tree-level amplitude 
(\ref{eq:defa1}); the sum over the two parities $(e,o)$ and $(o,e)$ of 
the cut lines reproduces the incoherent sum (\ref{eq:Asquared}).

%%%%%%%%%%%%%%%%%%%%%%%%%%%%%%%%%%%%%%%%%%%%%%%%%%%%%%%%%%%%%%%%%%%%%%%%%%%
%%%%%%%%%%%%%%%%%%%%%%%%%%%%%%%%%%%%%%%%%%%%%%%%%%%%%%%%%%%%%%%%%%%%%%%%%%%

%%%%%%%%%%%%%%%%%%%%%%%%%%%%%%%%%%%%%%%%%%%%%%%%%%%%%%%%%%%%%%%%%%%%%%%%%%%
%%%%%%%%%%%%%%%%%%%%%%%%%%% Conclusions %%%%%%%%%%%%%%%%%%%%%%%%%%%%%%%%%%%
%%%%%%%%%%%%%%%%%%%%%%%%%%%%%%%%%%%%%%%%%%%%%%%%%%%%%%%%%%%%%%%%%%%%%%%%%%%
%%%%%%%%%%%%%%%%%%%%%%%%%%%%%%%%%%%%%%%%%%%%%%%%%%%%%%%%%%%%%%%%%%%%%%%%%%%
%%%%%%%%%%%%%%%%%%%%%%%%%%%%%%%%%%%%%%%%%%%%%%%%%%%%%%%%%%%%%%%%%%%%%%%%%%%
\section{Conclusions}\label{sec:conc}
We have computed the exclusive transition amplitudes from the vacuum to a
two-particle state for a massless real scalar field coupled to a time-dependent
Dirichlet surface in $d+1$ dimensions, working perturbatively in the surface
deformation $\psi$ within the interaction picture and taking the Dirichlet limit
$\lambda\to\infty$ at the end. The amplitudes were obtained explicitly up to
third order in $\psi$ (\ref{eq:eqa1}),~(\ref{eq:defa1}), which is all that is
needed for the decay probabilities up to fourth order.

At second order the spectral pair-creation probability
(\ref{eq:dPgenD}) is a convolution of $|\widetilde\psi|^2$ with two on-shell
weights; in the localized, small-source regime it produces a dipole-like
$\cos\theta$ angular pattern peaked along the normal to the surface
(Fig.~\ref{fig:angular}). For a single-frequency drive we obtained the rate
$\gamma(\omega)$ in closed form, both in the Dirichlet limit and at finite
$\lambda$, where a mild bimodal structure develops and the total rate saturates
toward its Dirichlet value near $\lambda\sim K_s$ (Fig.~\ref{fig:gamma_finite_lambda}). The total second-order probability was evaluated in
closed form for arbitrary dimension, Eq.~(\ref{eq:Pdfinal}), and shown to
reproduce \emph{exactly} twice the imaginary part of the effective action of the
companion paper~\cite{FG2024}. This is the
dynamical-Casimir realization of the optical theorem: the discontinuity of the
one-loop bubble equals the on-shell phase-space integral of $|{\mathcal A}^{(1)}|^2$.

At fourth order a two-pair channel opens, and the naive identification
$P^{(4)}=2\,{\rm Im}\,\Gamma^{(4)}$ no longer holds. Using the Gaussian
(squeezed-vacuum) structure of the $S$-matrix, dictated by the quadratic
coupling, we showed that the exact decay probability exponentiates as
$1-e^{-2\,{\rm Im}\,\Gamma}$ and that, at $O(\psi^4)$, the one- and two-pair
probabilities satisfy $p_1^{(4)}+p_2^{(4)}=2\,{\rm Im}\,\Gamma^{(4)}
-2({\rm Im}\,\Gamma^{(2)})^2$, Eq.~(\ref{eq:P4decay}). 

Several directions invite further work. First, the closed-form evaluation of the
fourth-order phase-space integrals, the analogs of the $I_d,c_d$
coefficients of Sec.~\ref{sec:consistency}, would turn (\ref{eq:P4decay})
into an explicit check of the $\zeta_d$ coefficients of the effective action.
Second, the finite-$\lambda$ spectra obtained here suggest that the imperfect
(non-Dirichlet) regime, relevant to realistic boundaries, deserves a systematic
treatment; the angular and spectral signatures we found are, in principle,
observable imprints of the geometry and dynamics of the radiating surface. Finally, extending the analysis to more singular potentials, such as those implementing Neumann boundary conditions within the same approach, may shed light on the nontrivial treatment of singularities in this context.

\end{document}